\documentclass[preprint,12pt,preprintnumbers,amsmath,amssymb,floatfix,endfloats*]{revtex4}

\usepackage{graphicx}
\usepackage{dcolumn}
\usepackage{bm}
\usepackage{amsfonts}

\DeclareMathAlphabet{\mathsfsl}{OT1}{cmr}{bx}{it}
\begin{document}
%
\title{Oil droplet behavior at a pore entrance in the presence of crossflow:
Implications for microfiltration of oil-water dispersions}
\author{Tohid Darvishzadeh$^{1}$, Volodymyr V. Tarabara$^{2}$, Nikolai V. Priezjev$^{1}$}
\affiliation{$^{1}$Department of Mechanical Engineering, Michigan
State University, East Lansing, Michigan 48824}
\affiliation{$^{2}$Department of Civil and Environmental
Engineering, Michigan State University, East Lansing, Michigan
48824}
\date{\today}
%
\begin{abstract}

The behavior of an oil droplet pinned at the entrance of a micropore
and subject to clossflow-induced shear is investigated numerically
by solving the Navier-Stokes equation.  We found that in the absence
of crossflow, the critical transmembrane pressure required to force
the droplet into the pore is in excellent agreement with a
theoretical prediction based on the Young-Laplace equation.  With
increasing shear rate, the critical pressure of permeation
increases, and at sufficiently high shear rates the oil droplet
breaks up into two segments. The results of numerical simulations
indicate that droplet breakup at the pore entrance is facilitated at
lower surface tension, higher oil-to-water viscosity ratio and
larger droplet size but is insensitive to the value of the contact
angle. Using simple force and torque balance arguments, an estimate
for the increase in critical pressure due to crossflow and the
breakup capillary number is obtained and validated for different
viscosity ratios, surface tension coefficients, contact angles, and
drop-to-pore size ratios.

\end{abstract}

\maketitle

\section{Introduction}

Understanding the dynamics of an oil droplet at a pore entrance is a
fascinating problem at the intersection of fluid mechanics and
interface science that is of importance in such natural and
engineering processes as extraction of oil from bedrock,
lubrication, aquifer smearing by non-aqueous phase liquids, and
sealing of plant leaf
stomata~\cite{Morgan70,Erdemir05,Lee01,Kaiser07}. Membrane-based
separation of liquid-liquid dispersions and emulsions is a salient
example of a technology where the knowledge of liquid droplet
behavior in the vicinity of a surface pore is critical for the
success of practical applications. Milk fractionation, produced
water treatment, and recovery of electrodeposition paint are
examples of specific processes used in food, petroleum, and
automotive industries where porous membranes are relied on to
separate emulsions~\cite{Brans04,Veil11,Baker13}.

The membrane separation technique can be particularly useful when
small droplets need to be removed from liquid-liquid dispersions or
emulsions because other commonly used technologies, such as
hydrocyclones and centrifugation-based systems, are either incapable
of removing droplets smaller than a certain critical size (e.g.,
$\sim20\,{\mu}\text{m}$ for hydrocyclones) or are expensive and have
insufficient throughput (e.g., centrifuges).    The early work by
the Wiesner group~\cite{Nazzal96} and others~\cite{Cumming00} on oil
droplet entry into a pore provided an estimate of the critical
pressure of permeation; however, the understanding of the entire
process of the droplet dynamics at a micropore entrance is still
lacking, especially with regard to the practically-relevant case of
crossflow systems where blocking filtration laws~\cite{Hermia82}
are, strictly speaking, not applicable. Crossflow membrane
microfiltration is used to separate emulsions by shearing droplets
of the dispersed phase away from the membrane surface and letting
the continuous phase pass through~\cite{Koltuniewicz95}.      In
contrast to the normal, or dead-end, mode of filtration, crossflow
microfiltration allows for higher permeate fluxes due to better
fouling control~\cite{Mueller97}. However, the accumulation of the
dispersed phase on the surface of the membrane and inside the pores,
i.e., fouling of the membrane, can eventually reduce efficiency of
the process to an unacceptably low level even in the presence of
crossflow.

Another important application that entails interaction of liquid
droplets with porous media is membrane emulsification, where
micron-sized droplets are produced by forcing a liquid stream
through membrane pores into a channel where another liquid is
flowing~\cite{Gijsbertsen04}.     The emerging droplets break when
the viscous forces exerted by crossflow above the membrane surface
are larger than surface tension forces~\cite{Walstra93}. Membrane
emulsification requires less energy and produces a more narrow
droplet size distribution~\cite{Vladisavl02,Christopher07} than
conventional methods such as ultrasound
emulsification~\cite{Abismaill99} and stirring
vessels~\cite{Karbstein95}.

In general, the studies of petroleum emulsions have been performed
at two different scales, namely, macroscopic or bulk scales and
mesoscopic or droplet scales~\cite{Schramm92,Bremond12}.  Early
research on membrane emulsification and microfiltration involved
bulk experiments aimed at determining averaged quantities and
formulating empirical relations~\cite{Joscelyne00}.    These studies
considered macroscopic parameters such as droplet size distribution,
dispersed phase concentration, and bulk properties such as permeate
flux~\cite{Lee84,Hong03}.     These empirical approaches were
adopted due to inherent complexity of two-phase systems produced by
bulk emulsification, where shear stresses are spatially
inhomogeneous and the size distribution of droplets is typically
very broad~\cite{Atencia05,Bremond12}.   However, with the
development of imaging techniques and numerical methods, the shape
of individual droplets during deformation and breakup could be more
precisely quantified for various flow types and material
parameters~\cite{Stone94,Puyvelde08}.

First studies of the droplet dynamics date back to 1930's, when
G.~I. Taylor systematically investigated the deformation and breakup
of a single droplet in a shear flow~\cite{Taylor32,Taylor34}. Since
then, many groups have examined this problem
theoretically~\cite{Cox69,Hinch80} and in
experiments~\cite{Rumscheidt61,Grace82,Bentley86}.     A number of
research groups have studied experimentally how a droplet pinned at
the entrance of an unconfined pore deforms when it is exposed to a
shear flow~\cite{Christopher07}.     Experiments have also been
performed to measure the size of a droplet after breakup as a
function of shear rate and viscosity ratio~\cite{Husny06,Xu05}.
Numerical simulations of the droplet deformation and breakup have
been carried out using various methods including boundary
integral~\cite{Rallison81}, Lattice Boltzmann~\cite{Inamuro04}, and
Finite Volume~\cite{Tryggvason01} methods.  These multiphase flow
simulations generally use an interface-capturing method to track the
fluid interfaces.  Among other front-tracking methods, the Volume of
Fluid method simply defines the fluid-fluid interface through a
volume fraction function, which is updated based on the velocity
field obtained through the solution of the Navier-Stokes
equation~\cite{Brackbill92,Gueyffier99}.   The Volume of Fluid
method is mass-preserving, it is easily extendable to
three-dimensions, and it does not require special treatment to
capture topological changes~\cite{Scardovelli99}.

The drag force and torque on droplets or particles attached to a
solid substrate and subject to flow-induced shear stress depend on
their shape and the shear rate. Originally, O'Neill derived an exact
solution for the Stokes flow over a spherical particle on a solid
surface~\cite{Oneill68}. Later, Price computed the drag force on a
hemispherical bump on a solid surfaces under linear shear
flow~\cite{Price85}. Subsequently, Pozrikidis extended Price's work
to study the case of a spherical bump with an arbitrary angle using
the boundary integral method~\cite{Pozrikidis97}.   More recently,
Sugiyama and Sbragaglia~\cite{Sugiyama08} varied the viscosity ratio
to include values other than infinity (the only value considered by
Price~\cite{Price85}) and found an exact solution for the flow over
a hemispherical droplet attached to a solid surface. Assuming that
the droplet is pinned to the surface, an estimate for the drag
force, torque, and the deformation angle as a function of the
viscosity ratio was obtained analytically~\cite{Sugiyama08}. Also,
Dimitrakopoulos showed that the deformation and orientation of
droplets attached to solid surfaces under linear shear flow depend
on the contact angle, viscosity ratio, and contact angle
hysteresis~\cite{Dimitra07}.

More recently, Darvishzadeh and Priezjev~\cite{Darvishzadeh12}
studied numerically the entry dynamics of nonwetting oil droplets
into circular pores as a function of the transmembrane pressure and
crossflow velocity.  It was demonstrated that in the presence of
crossflow above the membrane surface, the oil droplets can be either
rejected by the membrane, permeate into a pore, or breakup at the
pore entrance. In particular, it was found that the critical
pressure of permeation increases monotonically with increasing shear
rate, indicating optimal operating conditions for the enhanced
microfiltration process.    However, the numerical simulations were
performed only for one specific set of parameters, namely, viscosity
ratio, contact angle, surface tension coefficient, and
droplet-to-pore size ratio. One of the goals of the present study is
to investigate the droplet dynamics in a wide range of material
parameters and shear rates.

In this paper, we examine the influence of physicochemical
parameters such as surface tension, oil-to-water viscosity ratio,
droplet size, and contact angle on the critical pressure of
permeation of an oil droplet into a membrane pore.    In the absence
of crossflow, our numerical simulations confirm analytical
predictions for the critical pressure of permeation based on the
Young-Laplace equation.  We find that when the crossflow is present
above the membrane surface, the critical pressure increases, and the
droplet deforms and eventually breaks up when the shear rate is
sufficiently high. Analytical predictions for the breakup capillary
number and the increase in critical permeation pressure due to
crossflow are compared with the results of numerical simulations
based on the Volume of Fluid method.

The rest of the paper is structured as follows.  In the next
section, the details of numerical simulations and a novel procedure
for computing the critical pressure of permeation are described.  In
Section\,\ref{sec:Results}, the summary of analytical predictions
for the critical pressure based on the Young--Laplace equation is
presented, and the effects of confinement, viscosity ratio, surface
tension, contact angle, and droplet size on the critical
transmembrane pressure and breakup are studied.   Conclusions are
provided in the last section.

\section{Details of numerical simulations}
\label{sec:Model}

\begin{figure}[t]
\includegraphics[width=13.0cm,angle=0]{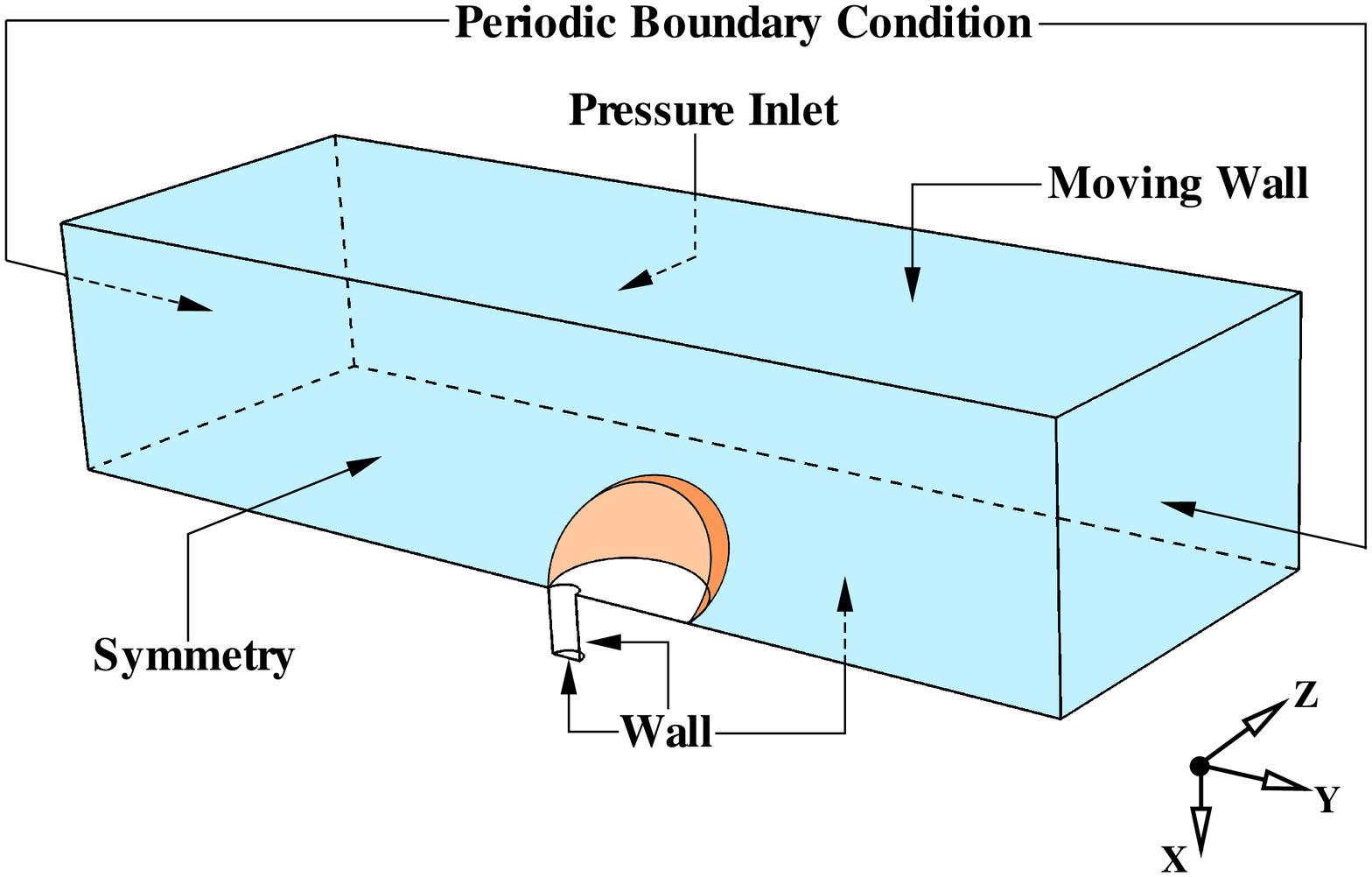}
\caption{Schematic representation of the oil droplet residing at the
pore entrance in a rectangular channel with the corresponding
boundary conditions.   The width and length of the computational
domain are fixed to $12\,{\mu}\text{m}$ and $36\,{\mu}\text{m}$,
respectively.  Symmetry boundary conditions are used in the
$\hat{z}$ direction.  The system dimensions are not drawn to scale.
} \label{fig:bc_schem}
\end{figure}

Three-dimensional numerical simulations were carried out using the
commercial software ANSYS FLUENT~\cite{fluent}.  The FLUENT flow
solver utilizes a control volume approach, while the Volume of Fluid
(VOF) method is implemented for the interface tracking in multiphase
flows.     In the VOF method, every computational cell contains a
certain amount of each phase specified by the volume fraction. For
two-phase flows, the volume fractions of $1$ and $0$ describe a
computational cell occupied entirely by one of the phases, while any
value in between corresponds to a cell that contains an interface
between the two phases~\cite{Hirt81}. In our simulations, GAMBIT was
employed to generate the mesh.    In order to increase the
simulation efficiency, we generated a hybrid mesh that consists of
fine hexagonal meshes in a part of the channel that contains the
droplet and coarse tetrahedral meshes in the rest of the channel. A
user-defined function was used to initialize the droplet shape and
to adjust the velocity of the top wall that induced shear flow in
the channel, as shown schematically in Fig.\,\ref{fig:bc_schem}.

\begin{figure}[t]
\includegraphics[width=14.0cm,angle=0]{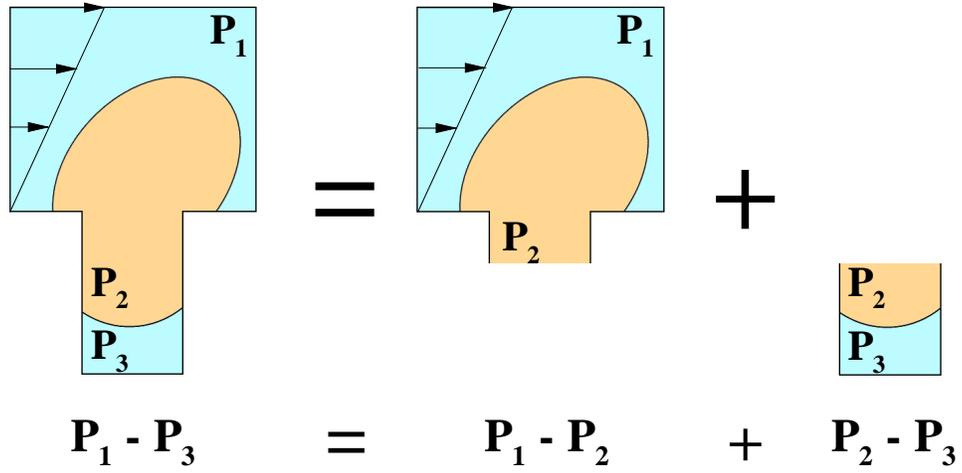}
\caption{Schematic of the droplet cross-sectional profile at the
membrane pore.    The critical pressure of permeation
($P_{1}-P_{3}$) is calculated in three steps: (1) the pressure jump
across the static interface ($P_{2}-P_{3}$) is calculated from the
Young--Laplace equation, (2) the pressure jump across the dynamic
interface ($P_{1}-P_{2}$) is computed numerically, and (3) the
pressure jumps from steps 1 and 2 are added. }
\label{fig:schematic_split}
\end{figure}

As we recently showed, the dynamics of the oil-water interface
inside the pore slows down significantly when the transmembrane
pressure becomes close to the critical pressure of
permeation~\cite{Darvishzadeh12}. Hence, the interface inside the
pore is nearly static and the pressure jump across the spherical
interface is given by the Young-Laplace equation.  However,
numerical simulations are required to resolve accurately the
velocity field, pressure, and shape of the deformed droplet above
the pore entrance. In the present study, we propose a novel
numerical procedure to compute the critical pressure of droplet
permeation and breakup, as illustrated in
Fig.\,\ref{fig:schematic_split}.  First, the pressure jump across
the static interface inside the pore is calculated using the
Young-Laplace equation.    Second, we simulate the oil droplet in
the presence of steady shear flow when the droplet covers the pore
entrance completely and the oil phase partly fills the pore.   In
the computational setup, the pore exit is closed to prevent the mass
flux and to keep the droplet at the pore entrance.   The difference
in pressure across the deformed oil-water interface with respect to
the inlet pressure is measured in the oil phase at the bottom of the
pore (see Fig.\,\ref{fig:schematic_split}).    The critical pressure
of permeation is then found by adding the pressure differences from
the first and second steps.   In the previous
study~\cite{Darvishzadeh12}, the critical pressure of permeation at
a given shear rate was determined iteratively by testing several
transmembrane pressures close to the critical pressure.  Using the
novel approach, we were able to reproduce our previous
results~\cite{Darvishzadeh12} faster and with higher accuracy.
Moreover, this numerical procedure was automated to detect the
critical pressure while increasing shear rate quasi-steadily, so
that less post-processing is required.

The solution of the Navier-Stokes equations for the flow over the
membrane surface requires specification of the appropriate boundary
conditions.      As shown in Fig.\,\ref{fig:bc_schem}, there are
four types of boundary conditions used in the computational domain.
The membrane surface is modeled as a no-slip boundary.    A moving
``wall" boundary condition is applied at the top surface of the
channel to induce shear flow between the moving top wall and the
stationary membrane surface. The bottom of the pore is also
described by the ``wall" boundary condition to prevent the mass flux
and to keep the oil droplet pinned at the pore entrance. Periodic
boundary conditions are imposed at the upstream and downstream
entries of the channel.     On the lateral side of the channel in
the $(Z+)$ direction, a pressure-inlet boundary condition is applied
to allow mass transfer, and to ensure that the reference pressure is
fixed. Finally, a ``symmetry" condition is implemented and only half
of the computational domain is simulated to reduce computational
efforts.   We performed test simulations with an oil droplet
$r_{d}=2\,{\mu}\text{m}$ exposed to shear flow and found that the
local velocity profiles at the upstream, downstream, and the lateral
sides remained linear when the width and length of the computational
domain were fixed to $12\,{\mu}\text{m}$ and $36\,{\mu}\text{m}$,
respectively.   These values were used throughout the study.    The
effect of confinement in the direction normal to the membrane
surface on the droplet deformation and breakup will be investigated
separately in the subsection\,\ref{subsec:confinement}.

The interface between two phases is described by a scalar variable,
known as the volume fraction $\alpha$, which is convected by the
flow at every iteration via the solution of the transport equation
as follows:
\begin{equation}
\frac{\partial{\alpha}}{{\partial}{t}}+{\nabla}{\cdot}\,({\alpha}{\textbf{V}})=\,0,
\label{eq:continuity}
\end{equation}
where $\textbf{V}$ is the three-dimensional velocity vector.    The
time dependence of the volume fraction is determined by the velocity
field near the interface.    Next, since the cells containing the
interface include both phases, the material properties are averaged
in each cell; for instance, the volume-fraction-averaged density is
computed as follows:
\begin{equation}
\rho = \alpha\,\rho_{2}+(1-\alpha)\,\rho_{1}.
\label{eq:compatibility}
\end{equation}
Using the averaged values of viscosity and density, the following
momentum equation is solved:
\begin{equation}
\frac{\partial}{{\partial}{t}}({\rho}{\textbf{V}})+\nabla\cdot(\rho{\textbf{V}}{\textbf{V}})
=-\nabla{p}+\nabla\cdot[\mu(\nabla{\textbf{V}}+\nabla{\textbf{V}}^{T})]+\rho\,\textbf{g}+\textbf{F},
\label{eq:momentum}
\end{equation}
where $\textbf{V}$ is the velocity vector shared between two phases,
$\textbf{g}$ is the gravitational acceleration, and $\textbf{F}$ is
the surface tension force per unit volume, which is given by
\begin{equation}
\textbf{F} =
\sigma\,\frac{\rho\,\kappa\nabla\alpha}{\frac{1}{2}(\rho_{1}+\rho_{2})},
\label{eq:surface_tension}
\end{equation}
where $\sigma$ is the surface tension coefficient and $\kappa$ is
the curvature of the oil-water interface, which in turn is defined
as
\begin{equation}
\kappa=\,\frac{1}{|\textbf{n}|}\Big[\Big(\frac{\textbf{n}}{|\textbf{n}|}\cdot\,\nabla\Big)|\textbf{n}|-
(\nabla\cdot\,\textbf{n})\Big], \label{eq:curvature}
\end{equation}
where $\textbf{n}$ is the vector normal to the interface.    The
surface tension force given by Eq.\,(\ref{eq:surface_tension}) is
nonzero only at the interface and it acts in the direction normal to
the interface ($\textbf{n}=\nabla\alpha$).   Segments with higher
interface curvature produce larger surface tension forces and tend
to smooth out the interface~\cite{Gerlach06}.     The orientation of
the interface at the wall is specified by the contact angle.    The
unit normal for a cell containing the interface at the wall is
computed as follows:
\begin{equation}
\textbf{n}_{i}=\,\textbf{n}_{w}\,\textrm{cos}\,\theta+\,\textbf{n}_{t}\,\textrm{sin}\,\theta,
\label{eq:contact_angle}
\end{equation}
where $\textbf{n}_{w}$ and $\textbf{n}_{t}$ are the unit vectors
normal to the wall and normal to the contact line at the wall,
respectively.   The angle $\theta$ is the static contact angle
measured in the dispersed phase~\cite{Brackbill92}.

A SIMPLE method was utilized for the pressure-velocity decoupling. A
second order upwind scheme was used for discretization of the
momentum equation and a staggered mesh with central differencing was
used for the pressure equation.   Piecewise Linear Interface
Reconstruction (PLIC) algorithm was employed to reconstruct the
interface in each cell~\cite{Rider98}. The continuum surface force
model of Brackbill\,\textit{et al.}~\cite{Brackbill92} was used to
compute the surface tension force.

An accurate computation of the pressure and velocity fields for
problems involving fluid interfaces requires a precise estimate of
the interfacial curvature.    It is well known that discrete
formulation of an interface produces a loss of accuracy in regions
of high curvature and, therefore, requires a sufficiently fine mesh.
The numerical simulations were performed using the mesh size of
$0.1\,{\mu}\text{m}$, which corresponds to $32$ mesh cells along the
perimeter of the membrane pore.    To ensure that the mesh
resolution is sufficiently high, we performed simulations at
different shear rates using $2$ and $4$ times finer meshes and found
that the resulting refinements in the final position of the droplet
interface and the values of the critical permeation pressure were
negligible.    The total volume of the oil phase inside the pore and
above the membrane surface was used to calculate the droplet radius.
Unless otherwise specified, the following parameters were used
throughout the study: the pore radius is $r_{p}=0.5\,{\mu}\text{m}$,
the droplet radius is $r_{d}=2\,{\mu}\text{m}$, the contact angle is
$\theta=135^{\circ}$, and the surface tension coefficient is
$\sigma=19.1\,\text{mN/m}$.

\section{Results}
\label{sec:Results}

\subsection{The critical pressure of permeation and the breakup capillary number}
\label{subsec:analytical}

The pressure jump across a static interface between two immiscible
fluids can be determined from the Young--Laplace equation as a
product of the interfacial tension coefficient and the mean
curvature of the interface or ${\Delta}P = 2\,\sigma\,\kappa$. For a
pore of arbitrary cross-section, the mean curvature of the interface
is given by
\begin{equation}
\kappa =\frac{C_{p}\,\textrm{cos}\,{\theta}}{2\,A_{p}},
\label{eq:mean_curv}
\end{equation}
where $C_{p}$ and $A_{p}$ are the cross-sectional circumference and
area of the pore, respectively~\cite{Concus90}.   Therefore, the
critical pressure of permeation of a liquid film into a pore of
arbitrary cross-section is given by
\begin{equation}
P_{cr} = \frac{\sigma\,C_p\,\textrm{cos}\,\theta}{A_p}.
\label{eq:pcr_arbitrary}
\end{equation}
In our recent study~\cite{Darvishzadeh12}, the theoretical
prediction for the critical permeation pressure,
Eq.\,(\ref{eq:pcr_arbitrary}), was validated numerically for oil
films on a membrane surface with rectangular, elliptical, and
circular pores.

In the case of a liquid droplet blocking a membrane pore, the
critical pressure of permeation, Eq.\,(\ref{eq:pcr_arbitrary}), has
to be adjusted to account for the finite size of the droplet. It was
previously shown~\cite{Nazzal96,Cumming00} that the critical
pressure for an oil droplet of radius $r_{d}$ to enter a circular
pore of radius $r_{d}$ is given by
\begin{equation}
P_{cr} =
\frac{2\,\sigma\,\textrm{cos}\,\theta}{r_{p}}\,\,\sqrt[3]{1-\frac{2+3\,\textrm{cos}\,\theta-
\textrm{cos}^3\theta}{4\,(r_{d}/r_{p})^3\,\textrm{cos}^3\theta-(2-3\,\textrm{sin}\theta+\textrm{sin}^3\theta)}}.
\label{eq:pcritfull}
\end{equation}
We showed earlier that the analytical prediction for the critical
pressure given by Eq.\,(\ref{eq:pcritfull}) agrees well with the
results of numerical simulations for an oil droplet at the pore
entrance in the absence of crossflow~\cite{Darvishzadeh12}.    In
the presence of crossflow, however, Eq.\,(\ref{eq:pcritfull}) in not
valid as the shear flow deforms the droplet rendering its interface
above the membrane surface non-spherical~\cite{Darvishzadeh12}.
Furthermore, numerical simulations have shown that the critical
pressure of permeation increases with increasing crossflow velocity
up to a certain value, above which the droplet breaks
up~\cite{Darvishzadeh12}.   Hence, the phase diagram was determined
for the droplet rejection, permeation, and breakup depending on the
transmembrane pressure and shear rate~\cite{Darvishzadeh12}.   In
the present study, the critical permeation pressure is determined
more accurately and its dependence on shear rate is studied
numerically for a range of material properties and geometrical
parameters.

In the presence of crossflow above the membrane surface, an oil
droplet breaks up when viscous stresses over the droplet surface
exposed to the flow become larger than capillary stresses at the
interface of the droplet near the membrane pore.    Therefore, at
the moment of breakup, the drag force in the flow direction is
balanced by the capillary force at the droplet interface around the
pore
\begin{equation}
D \approx F_{\sigma}. \label{eq:force_balance}
\end{equation}
Neglecting the contact angle dependence, $F_{\sigma} \propto
\sigma\,r_{p}$ is the interfacial force acting in the direction
opposite to the flow at the droplet interface near the pore
entrance. The drag force generated by a linear shear flow on a
spherical droplet attached to a solid surface is given by
\begin{equation}
D \propto f_{D}(\lambda)\,\mu\,\dot{\gamma}\,r_{d}^2,
\label{eq:drag}
\end{equation}
where $\mu$ is the viscosity of the continuous phase, $\dot{\gamma}$
is the shear rate, and $r_{d}$ is the radius of the
droplet~\cite{Zapryanov98,Sugiyama08}.   The coefficient
$f_{D}(\lambda)$ is a function of the viscosity ratio
$\lambda=\mu_{oil}/\mu_{water}$ and it depends on the shape of the
droplet above the surface.  Sugiyama and
Sbragaglia~\cite{Sugiyama08} have estimated this function
analytically for a hemispherical droplet ($\theta=90^{\circ}$)
attached to a solid surface
\begin{equation}
f_{D}(\lambda) \approx \frac{2+4.510\,\lambda}{1+1.048\,\lambda}.          
\label{eq:f_lambda_force}
\end{equation}
By plugging Eq.\,(\ref{eq:drag}) into Eq.\,(\ref{eq:force_balance})
and introducing $\bar{r}=\,r_{d}/\,r_{p}$, the critical capillary
number for breakup of a droplet on a pore can be expressed as
follows:
\begin{equation}
Ca_{cr} \propto \frac{1}{f_{D}(\lambda)\,\bar{r}}, \label{eq:ca_cr}
\end{equation}
where the capillary number is defined as
$Ca={\mu_w}\dot{\gamma}r_{d}/\sigma$.

The difference in pressure inside the pore in the presence of flow
and at zero shear rate can be estimated from the torque generated by
the shear flow on the droplet surface.   The torque around the
center of the droplet projected on the membrane surface is given by
\begin{equation}
T \propto f_{T}(\lambda)\,\mu\,\dot{\gamma}\,r_{d}^3,
\label{eq:torque}
\end{equation}
It was previously shown~\cite{Sugiyama08} that for a hemispherical
droplet on a solid surface, $f_{T}(\lambda)$ is a function of the
viscosity ratio
\begin{equation}
f_{T}(\lambda) \approx \frac{2.188\,\lambda}{1+0.896\,\lambda}.
\label{eq:f_lambda_torque}
\end{equation}
Hence, the balance of the torque due to shear flow above the
membrane surface [given by Eq.\,(\ref{eq:torque})] and the torque
arising from the pressure difference,
$(P_{cr}-P_{cr_{0}})\,A_{p}\,{r_{d}}$, can be reformulated in terms
of the capillary number and drop-to-pore size ratio as follows:
\begin{equation}
P_{cr}-P_{cr_{0}} \propto
\frac{f_{T}(\lambda)\,\sigma\,\bar{r}\,Ca}{r_{p}},
\label{eq:pressure_cr}
\end{equation}
where $P_{cr_{0}}$ is the critical permeation pressure in the
absence of crossflow.

In what follows, we consider the effects of confinement, viscosity
ratio, surface tension, contact angle, and droplet size on the
critical pressure of permeation and breakup using numerical
simulations and analytical predictions of Eq.\,(\ref{eq:ca_cr}) and
Eq.\,(\ref{eq:pressure_cr}).

\subsection{The effect of confinement on droplet deformation and breakup}
\label{subsec:confinement}

In practical applications, the dimensions of a crossflow channel of
a microfiltration system are much larger than the typical size of
emulsion droplets so that the velocity profile over the distance of
about $r_d$ from the membrane surface can be approximated as linear.
To more closely simulate this condition in our computational setup,
the shear flow above the membrane surface was induced by moving the
upper wall of the crossflow channel (Fig.\,\ref{fig:bc_schem}).   To
understand how the finite size of the channel affects droplet
dynamics at the membrane surface, we studied the influence of the
channel height on the droplet behavior.    The confinement ratio is
defined as the ratio of the height of the droplet residing on the
pore at zero shear rate $H_{d}$ (i.e., the height of a spherical cap
above the membrane surface) to the channel height $H_{ch}$.    It is
important to note that the degree of confinement is varied only in
the direction normal to the membrane surface and the computational
domain is chosen to be wide enough for the lateral confinement
effects to be negligible (see Section\,\ref{sec:Model}).

\begin{figure}[t]
\includegraphics[width=12.0cm,angle=0]{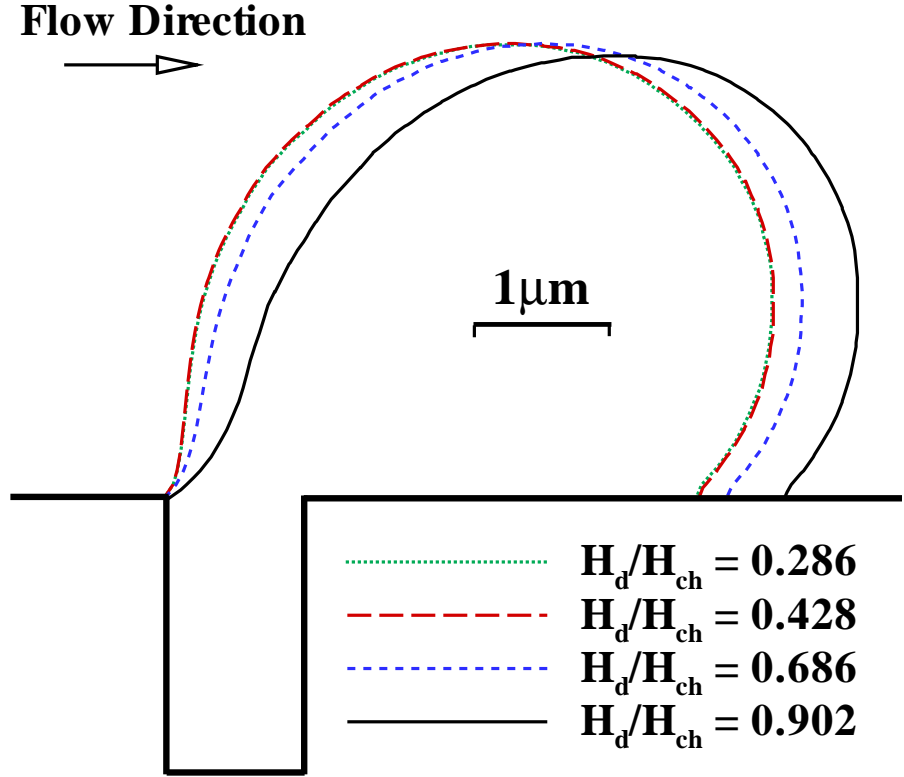}
\caption{The cross-sectional profiles of oil droplets in steady
shear flow for the indicated confinement ratios when the capillary
number is $Ca=0.021$.   The droplet radius is
$r_{d}=2\,{\mu}\text{m}$, the pore radius is
$r_{p}=0.5\,{\mu}\text{m}$, the contact angle is
$\theta=135^{\circ}$, the surface tension coefficient is
$\sigma=19.1\,\text{mN/m}$, and the viscosity ratio is $\lambda=1$.
} \label{fig:confinement_prof}
\end{figure}

We performed numerical simulations of an oil droplet with radius
$r_d=2\,{\mu}\text{m}$ in steady-state shear flow for the channel
heights $3.8\,{\mu}\text{m} \leqslant H_{ch} \leqslant
12.0\,{\mu}\text{m}$.    Figure\,\ref{fig:confinement_prof}
illustrates the effect of confinement on the shape of the droplet
residing on a $r_{p}=0.5\,{\mu}\text{m}$ pore when the capillary
number is $Ca={\mu_w}\dot{\gamma}r_{d}/\sigma=0.021$.   The height
of the droplet above the membrane surface in the absence of flow is
approximately $3.43\,{\mu}\text{m}$.    It can be observed from
Fig.\,\ref{fig:confinement_prof} that highly confined droplets
become more elongated in the direction of flow than droplets with
lower confinement ratios, which is in agreement with the results of
previous simulations~\cite{Renardy07}.   When a droplet is highly
confined, the distance between the upper moving wall and the top of
the droplet is relatively small.    As a result, the effective shear
rate at the surface of the droplet is higher and the droplet
undergoes larger deformation.    Furthermore, the cross-sectional
profiles for the confinement ratios of $0.428$ and $0.286$ are
nearly identical, indicating that the flow around the droplet is not
affected by the upper wall when $H_d/H_{ch} \lesssim 0.428$ and the
capillary number is fixed.

\begin{figure}[t]
\includegraphics[width=12.0cm,angle=0]{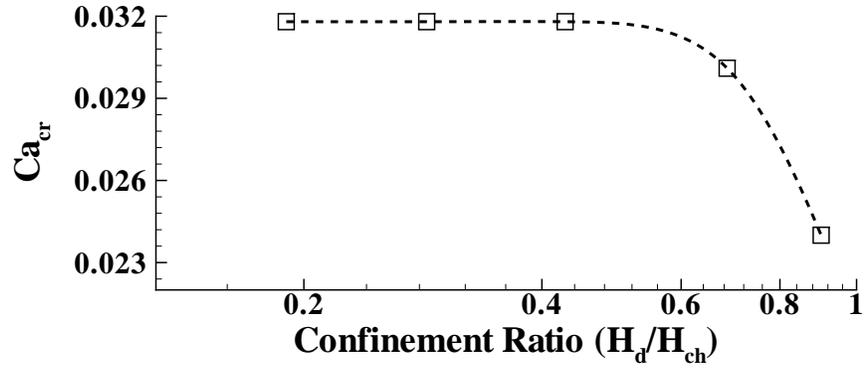}
\caption{The critical (breakup) capillary number as a function of
the confinement ratio $H_d/H_{ch}$.    Other parameters are the same
as in Fig.\,\ref{fig:confinement_prof}.  }
\label{fig:confinement_breakup}
\end{figure}

Figure\,\ref{fig:confinement_breakup} shows the variation of the
critical capillary number (right before breakup) as a function of
the confinement ratio for the the same material parameters as in
Fig.\,\ref{fig:confinement_prof}.    These results indicate that
highly confined droplets breakup at lower capillary numbers, and,
when the confinement ratio is smaller than about $0.5$, the breakup
capillary number remains nearly constant.  For the rest of the
study, the channel height was fixed to $8\,{\mu}\text{m}$, which
corresponds to the confinement ratio of $0.428$ for a droplet with
radius $r_{d}=\,2\,{\mu}\text{m}$.   For the results presented in
the subsection\,\ref{subsec:droplet_size}, the channel height was
scaled appropriately to retain the same confinement ratio for larger
droplets.

\subsection{The effect of viscosity ratio on the critical transmembrane pressure}
\label{subsec:viscosity}

The ratio of viscosities of the dispersed and continuous phases is
an important factor that determines the magnitude of viscous
stresses at the interface between the two phases.     For a small
droplet at low Reynolds numbers, the viscous stresses are primarily
counterbalanced by interfacial tension stresses.    In a shear flow,
viscous stresses tend to distort the surface of a droplet, while
interfacial stresses assist in retaining its initial spherical
shape.    The competition between the two stresses determines the
breakup criterion, deformation, and orientation of the
droplet~\cite{Stone94,Rallison84}.    In this subsection, we
investigate numerically the effect of viscosity ratio on the droplet
deformation and breakup at the entrance of the membrane pore.

\begin{figure}[t]
\includegraphics[width=12.0cm,angle=0]{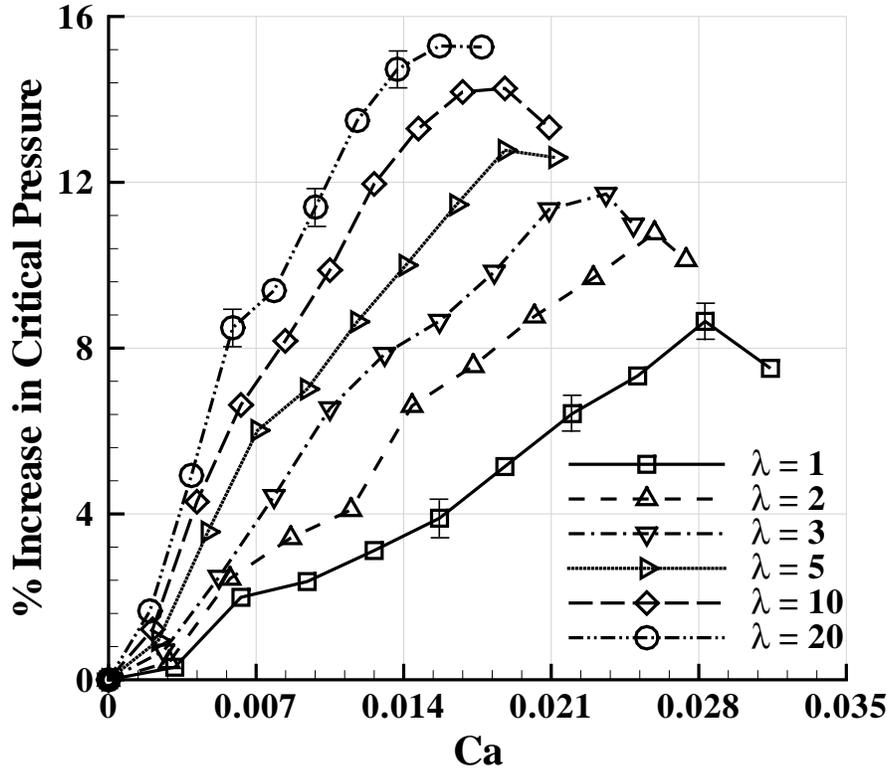}
\caption{The percent increase in critical pressure of permeation as
a function of the capillary number
$Ca={\mu_w}\dot{\gamma}r_{d}/\sigma$ for the indicated viscosity
ratios $\lambda=\mu_o/\mu_w$.  Typical error bars are shown on
selected data points.   For each value of $\lambda$, the data are
reported up to the critical capillary number above which droplets
break into two segments.   The droplet and pore radii are
$r_{d}=2\,{\mu}\text{m}$ and $r_{p}=0.5\,{\mu}\text{m}$,
respectively.     The contact angle is $\theta=135^{\circ}$ and the
surface tension coefficient is $\sigma=19.1\,\text{mN/m}$. }
\label{fig:vis_percent_inc}
\end{figure}

Figure\,\ref{fig:vis_percent_inc} shows the effect of the viscosity
ratio, $\lambda=\mu_o/\mu_w$, on the critical pressure of permeation
and breakup of an oil droplet on a membrane pore as a function of
the capillary number.     The percent increase in critical pressure
is defined with respect to the critical pressure in the absence of
crossflow $P_{cr_{0}}$, i.e.,
$(P_{cr}-P_{cr_{0}})/P_{cr_{0}}\times100\,\%$.   Keeping in mind
that $P_{cr_{0}}$ does not depend on $\lambda$, the results shown in
Fig.\,\ref{fig:vis_percent_inc} demonstrate that at a fixed $Ca$,
the critical pressure increases with increasing viscosity ratio,
which implies that higher viscosity droplets penetrate into the pore
at higher transmembrane pressures.    Specifically, the maximum
increase in critical pressure just before breakup is about $8\,\%$
for $\lambda=1$ and about $15\,\%$ for $\lambda=20$. Furthermore,
highly viscous droplets tend to break at lower shear rates because
of the larger torque generated by the shear flow [see
Eq.\,(\ref{eq:torque})].    As reported in
Fig.\,\ref{fig:vis_percent_inc}, the critical capillary number for
breakup varies from about $0.018$ for $\lambda=20$ to $0.032$ for
$\lambda=1$.    The practical implication of these results is that
in membrane emulsification processes the use of liquids with lower
viscosity ratios should be avoided as the droplets tend to break at
higher shear rates.

\begin{figure}[t]
\includegraphics[width=12.0cm,angle=0]{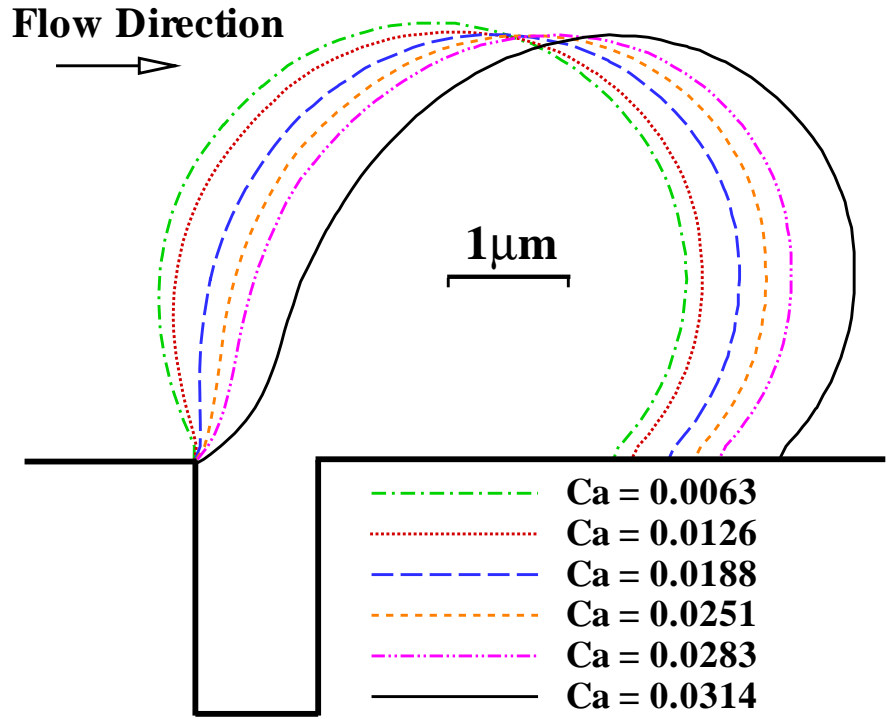}
\caption{The cross-sectional profiles of the oil droplet residing on
the circular pore with $r_{p}=0.5\,{\mu}\text{m}$ for the indicated
capillary numbers.    The viscosity ratio is $\lambda=1$.    Other
parameters are the same as in Fig.\,\ref{fig:vis_percent_inc}.    }
\label{fig:lambda1_profiles}
\end{figure}

Examples of cross-sectional profiles of the oil droplet in steady
shear flow are presented in Fig.\,\ref{fig:lambda1_profiles} for the
viscosity ratio $\lambda=1$.     At small capillary numbers, no
significant deformation occurs and the droplet retains its spherical
shape above the membrane surface.      As $Ca$ increases, a neck
forms at the pore entrance while the rest of the droplet remains
nearly spherical.     A closer look at the shapes of the droplet for
$Ca=0.0283$ and $0.0314$ in Fig.\,\ref{fig:lambda1_profiles} reveals
that with increasing shear flow, the neck gets thinner and the
droplet becomes more elongated in the direction of flow.  While the
torque due to the shear flow does not increase significantly, the
elongated shape of the droplet results in an effectively longer arm
for the torque due to pressure in the droplet along the flow
direction, and, thus, it leads to a lower critical permeation
pressure required to keep the droplet attached to the pore.   This
effect is observed in Fig.\,\ref{fig:vis_percent_inc} as the
critical pressure just before breakup decreases as a function of
$Ca$.

\begin{figure}[t]
\includegraphics[width=12.0cm,angle=0]{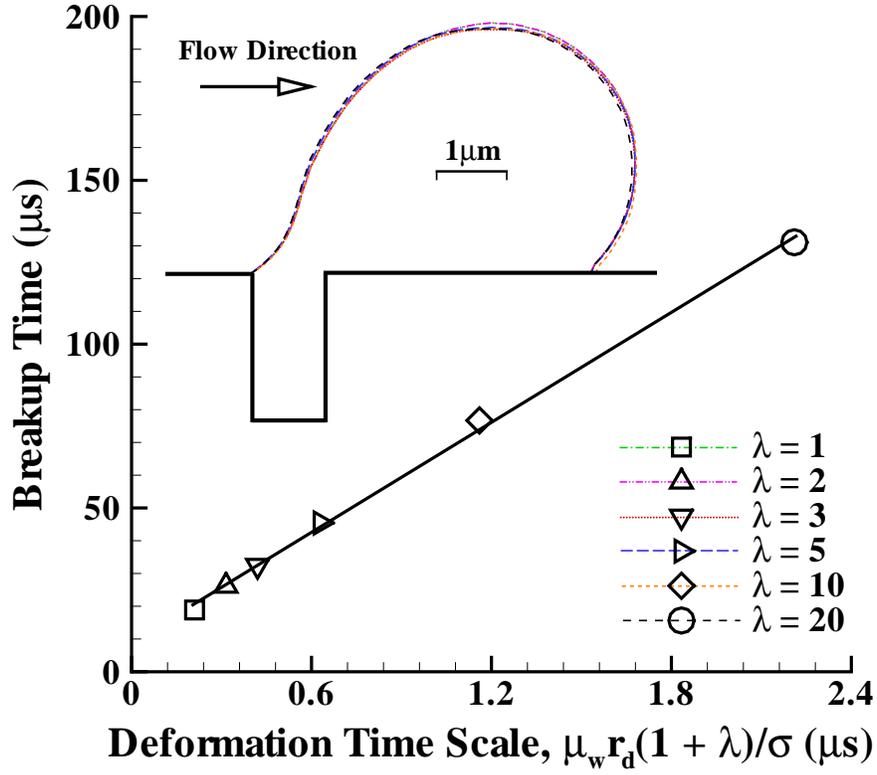}
\caption{The breakup time versus deformation time scale
$\mu_{w}r_{d}\,(1+\lambda)/\sigma$ for the tabulated values of the
viscosity ratio $\lambda=\mu_o/\mu_w$.   Other system parameters are
the same as in Fig.\,\ref{fig:vis_percent_inc}.    The straight line
is the best fit to the data.    The error bars for the breakup time
are about the symbol size.   The inset shows the droplet profiles
just before breakup for the same viscosity ratios.  }
\label{fig:vis_assembled}
\end{figure}

We next estimate the breakup time and compare it with the typical
deformation time of the droplet interface for different viscosity
ratios.    In our simulations, the upper wall velocity is increased
quasi-steadily and the spontaneous initiation of the breakup process
can be clearly detected by visual inspection of the droplet
interface near the pore entrance.    We then identify the moment
when a droplet breaks into two segments and compute the breakup
time.   The deformation time scale, defined by
$\mu_{w}r_{d}\,({1+\lambda})/\sigma$, is a measure of the typical
relaxation time of the droplet interface with respect to its
deformation at steady state~\cite{Dimitra07,Stone88}.      In
Fig.\,\ref{fig:vis_assembled}, the breakup time is plotted against
the deformation time scale for different viscosity ratios.    Notice
that the breakup time increases linearly with the deformation time
scale, which confirms that highly viscous droplets break up more
slowly.     The inset in Fig.\,\ref{fig:vis_assembled} displays the
droplet cross-sectional profiles just before breakup for the same
viscosity ratios.   It can be observed that the profiles nearly
overlap with each other, indicating that droplets with different
viscosities are deformed identically just before breakup.

\begin{figure}[t]
\includegraphics[width=12.0cm,angle=0]{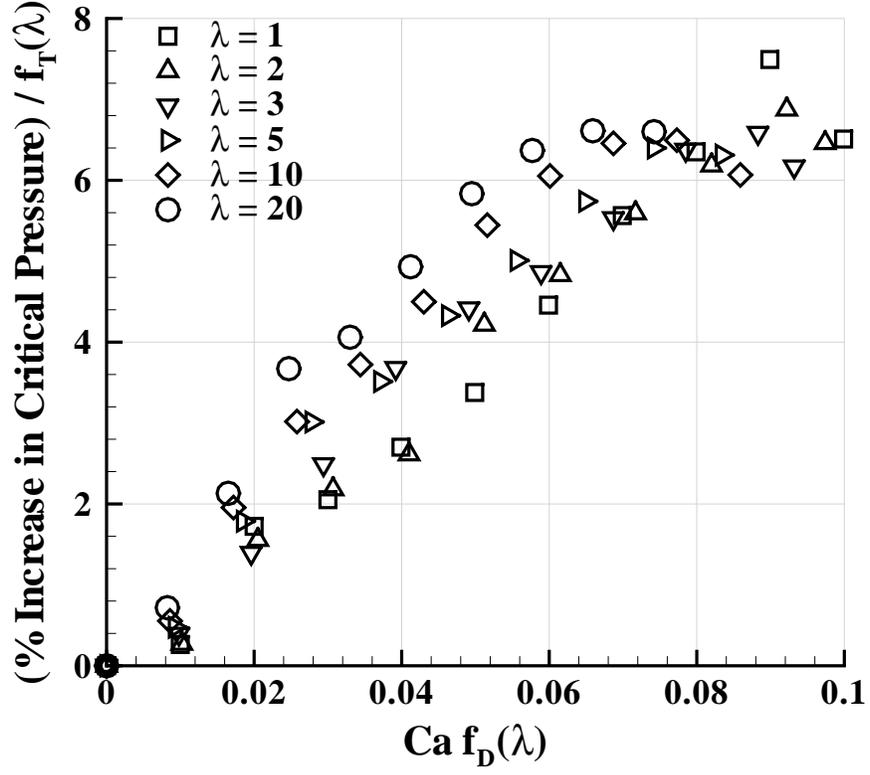}
\caption{The normalized percent increase in critical pressure of
permeation versus the modified capillary number $Ca\,f_{D}(\lambda)$
for the selected values of the viscosity ratio
$\lambda=\mu_o/\mu_w$. The functions $f_{D}(\lambda)$ and
$f_{T}(\lambda)$ are given by Eq.\,(\ref{eq:f_lambda_force}) and
Eq.\,(\ref{eq:f_lambda_torque}), respectively.  }
\label{fig:vis_percent_inc_mod}
\end{figure}

According to Eq.\,(\ref{eq:ca_cr}), the breakup capillary number
depends on the drop-to-pore size ratio and the viscosity ratio via
the function $f_{D}(\lambda)$.     Therefore, it is expected that
the product $Ca_{cr}\,f_{D}(\lambda)$ will be independent of
$\lambda$ and the appropriate dimensionless number for a constrained
viscous droplet in a shear flow is $Ca\,f_{D}(\lambda)$. Moreover,
based on Eq.\,(\ref{eq:pressure_cr}), the percent increase in the
critical pressure is independent of the viscosity ratio when it is
divided by $f_{T}(\lambda)$. Figure\,\ref{fig:vis_percent_inc_mod}
shows the same data as in Fig.\,\ref{fig:vis_percent_inc} but
replotted in terms of the normalized critical pressure and the
modified capillary number.   As is evident from
Fig.\,\ref{fig:vis_percent_inc_mod}, the data for different
viscosity ratios nearly collapse on the master curve.   It is seen
that droplets break at approximately the same value
$Ca\,f_{D}(\lambda) \approx 0.09$.    In practice, the increase in
critical pressure due to crossflow can be roughly estimated from the
master curve in Fig.\,\ref{fig:vis_percent_inc_mod} for any
viscosity ratio in the range $1 \leqslant \lambda \leqslant 20$.
Also, if $Ca\,f_{D}(\lambda) \gtrsim 0.09$, the oil droplets will
break near the pore entrance for any viscosity ratio.

\subsection{The effect of surface tension on the critical pressure of permeation}
\label{subsec:surface_tension}

\begin{figure}[t]
\includegraphics[width=12.0cm,angle=0]{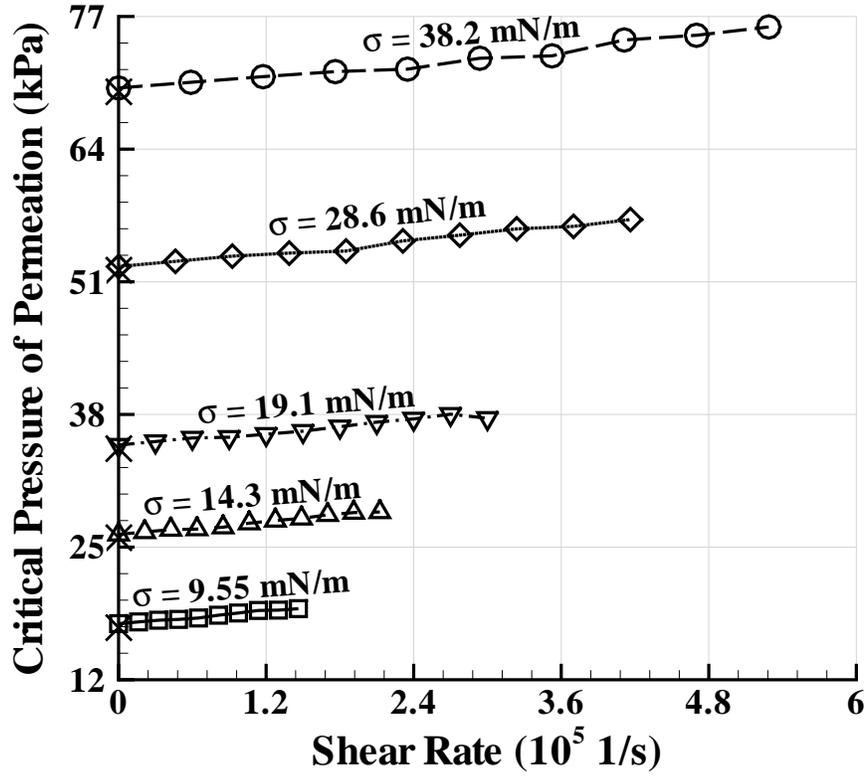}
\caption{The critical pressure of permeation as a function of shear
rate for the indicated surface tension coefficients. The symbols
($\times$) denote the analytical predictions of
Eq.\,(\ref{eq:pcritfull}). The droplet and pore radii are
$r_{d}=2\,{\mu}\text{m}$ and $r_{p}=0.5\,{\mu}\text{m}$,
respectively.   The viscosity ratio is $\lambda=1$ and the contact
angle is $\theta=135^{\circ}$.  } \label{fig:sigma_pcr}
\end{figure}

In this subsection, we investigate the influence of surface tension
on the critical permeation pressure, deformation and breakup of an
oil droplet residing at the pore entrance in the presence of
crossflow above the membrane surface.    Figure\,\ref{fig:sigma_pcr}
shows the critical pressure of permeation as a function of shear
rate for five values of the surface tension coefficient.    As
expected from Eq.\,(\ref{eq:pcritfull}), the critical pressure at
zero shear rate increases linearly with increasing surface tension
coefficient.     Note that oil droplets with higher surface tension
break up at higher shear rates because larger stresses are required
to deform the interface and cause breakup of the neck.     Also, the
difference between the critical pressure just before breakup and
$P_{cr_{0}}$ is larger at a higher surface tension; for example, it
is about 1.5\,kPa for $\sigma=9.55\,\text{mN/m}$ and 6\,kPa for
$\sigma=38.2\,\text{mN/m}$.       The results shown in
Fig.\,\ref{fig:sigma_pcr} suggest that crossflow microfiltration of
emulsion droplets with higher surface tension is more efficient
because higher transmembrane pressure can be applied and the droplet
breakup is less likely.

\begin{figure}[t]
\includegraphics[width=12.0cm,angle=0]{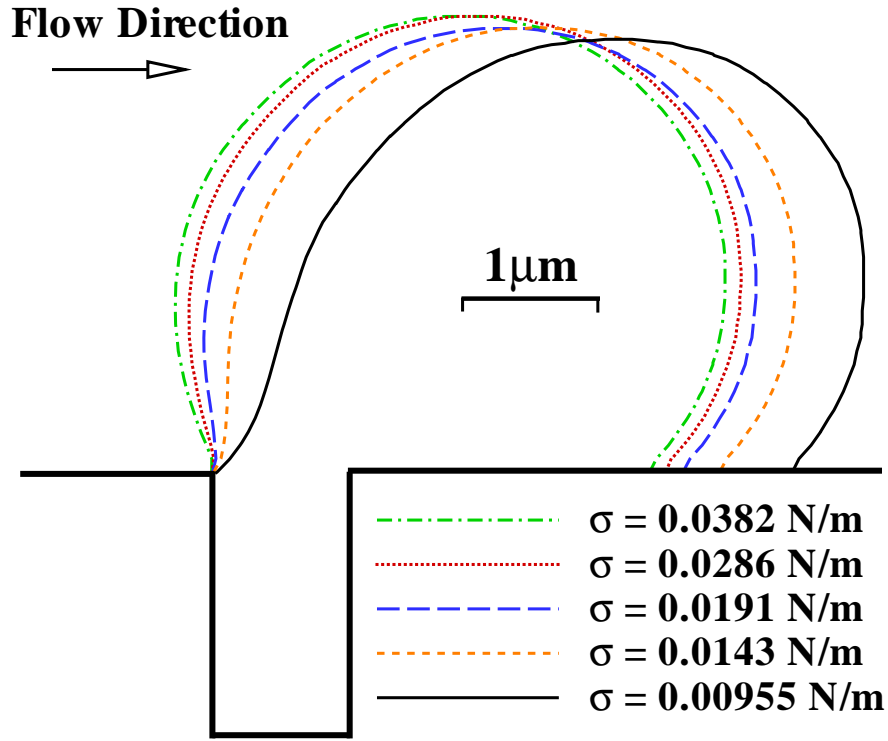}
\caption{The cross-sectional profiles of the oil droplet above the
circular pore for the listed values of the surface tension
coefficient.   In all cases, the shear rate is
$\dot{\gamma}=1.5\times10^{5}\,\text{s}^{-1}$.   Other parameters
are the same as in Fig.\,\ref{fig:sigma_pcr}. }
\label{fig:sigma_gammadot_profs}
\end{figure}

Examples of droplet cross-sectional profiles above the membrane pore
are presented in Fig.\,\ref{fig:sigma_gammadot_profs} for five
values of the surface tension coefficient.     These profiles are
extracted from the data reported in Fig.\,\ref{fig:sigma_pcr} at the
shear rate $\dot{\gamma}=1.5\times10^{5}\,\text{s}^{-1}$.  It can be
observed that oil droplets with lower surface tension become highly
deformed along the flow direction.   The elongation is especially
pronounced when the surface tension coefficient is small; for
$\sigma=9.55\,\text{mN/m}$ the droplet interface is deformed locally
near the pore entrance and the neck is formed.

\begin{figure}[t]
\includegraphics[width=12.0cm,angle=0]{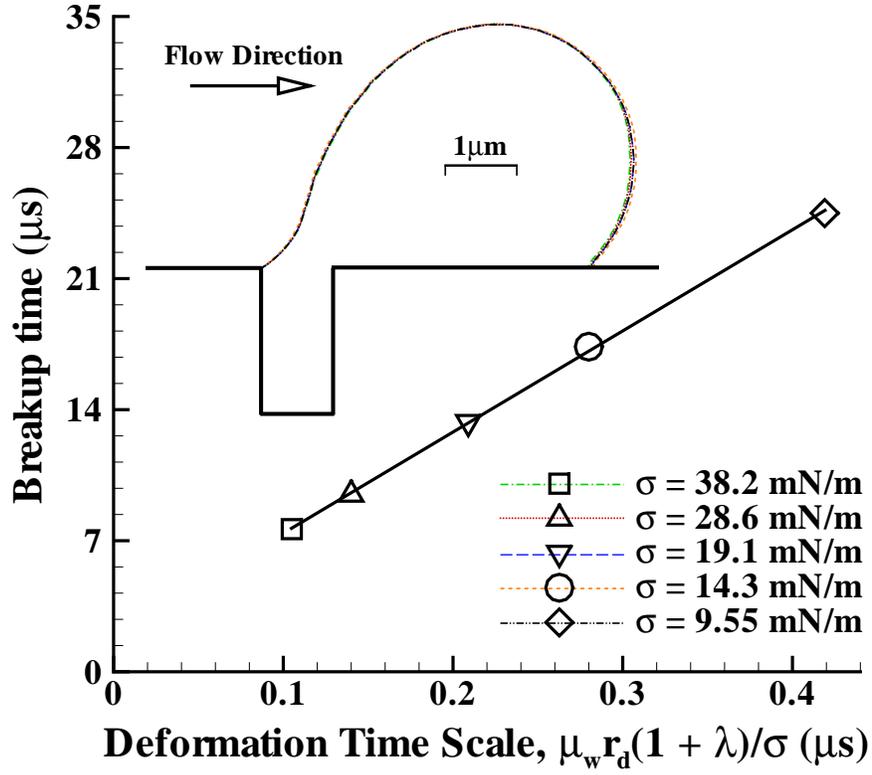}
\caption{The breakup time versus deformation time scale
$\mu_{w}r_{d}\,(1+\lambda)/\sigma$ for the surface tension
coefficients in the range from $9.55\,\text{mN/m}$ to
$38.2\,\text{mN/m}$.   Other parameters are the same as in
Fig.\,\ref{fig:sigma_pcr}.  The straight solid line is the best fit
to the data.   The cross-sectional profiles of the oil droplet just
before breakup are displayed in the inset. }
\label{fig:sigma_assembled}
\end{figure}

To further investigate the effect of surface tension on the droplet
breakup, we compare the breakup time and the deformation time scale
$\mu_{w}r_{d}\,(1+\lambda)/\sigma$.   The numerical results are
summarized in Fig.\,\ref{fig:sigma_assembled} for the same values of
the surface tension coefficient as in Fig.\,\ref{fig:sigma_pcr}.
Similar to the analysis in the previous subsection, the breakup time
was estimated from the time when a droplet becomes unstable under
quasi-steady perturbation till the formation of two separate
segments.    It can be observed in Fig.\,\ref{fig:sigma_assembled}
that the breakup time varies linearly with increasing deformation
time scale, which in turn indicates that the breakup time is
inversely proportional to the surface tension coefficient.   In
addition, the inset in Fig.\,\ref{fig:sigma_assembled} shows the
cross-sectional profiles of the droplet just before breakup for the
same surface tension coefficients.    Interestingly, the profiles
nearly coincide with each other, indicating that the droplet shape
at the moment of breakup is the same for any surface tension.

\begin{figure}[t]
\includegraphics[width=12.0cm,angle=0]{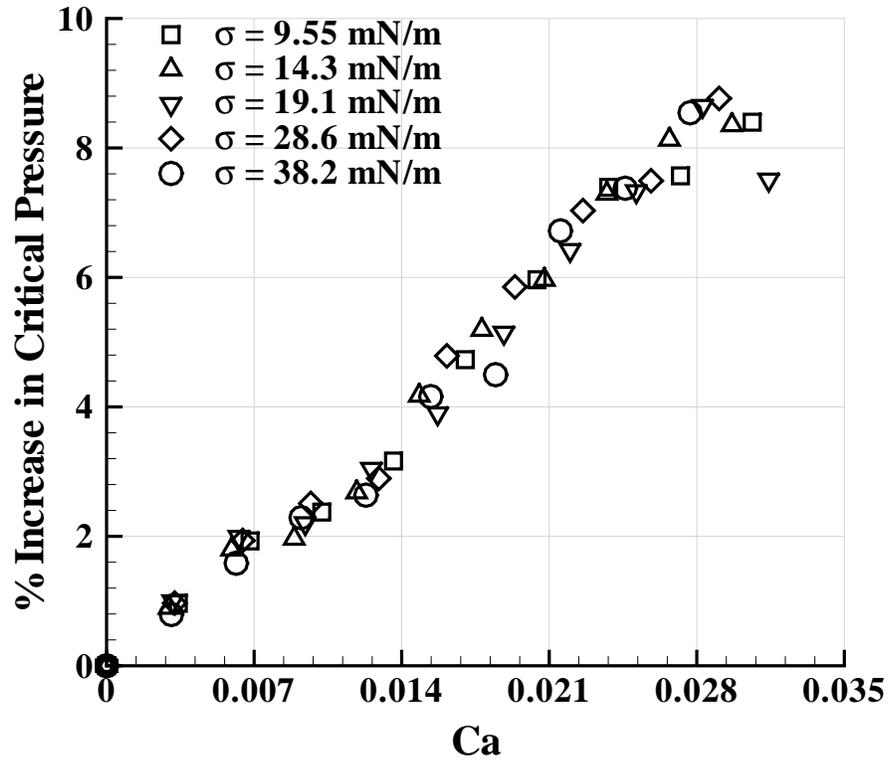}
\caption{The percent increase in critical pressure of permeation as
a function of the capillary number
$Ca={\mu_w}\dot{\gamma}r_{d}/\sigma$ for the selected values of the
surface tension coefficient.    The rest of the material parameters
are the same as in Fig.\,\ref{fig:sigma_pcr}. }
\label{fig:sigma_pcr_nondim}
\end{figure}

In order to present our results in a more general form, we replotted
the data from Fig.\,\ref{fig:sigma_pcr} in terms of the percent
increase in critical pressure,
$(P_{cr}-P_{cr_{0}})/P_{cr_{0}}\times100\,\%$, and the capillary
number in Fig.\,\ref{fig:sigma_pcr_nondim}. Note that in all cases,
the data collapse onto a master curve and breakup occurs at the same
relative pressure $(P_{cr}-P_{cr_{0}})/P_{cr_{0}}\approx 8\,\%$ and
$Ca_{cr}\approx0.03$, which indicates that the capillary number is
an appropriate dimensionless number to describe the droplet
deformation in shear flow with variable surface tension. These
results are not surprising, given that the breakup capillary number,
Eq.\,(\ref{eq:ca_cr}), does not depend on the surface tension
coefficient.    Moreover, the increase in critical pressure due to
crossflow, Eq.\,(\ref{eq:pressure_cr}), is proportional to $\sigma$
and $Ca$, and when it is divided by $P_{cr_{0}}$, which itself is a
linear function of $\sigma$ [see Eq.\,(\ref{eq:pcritfull})], the
percent increase in critical pressure becomes proportional to the
capillary number.    In practice, the master curve reported in
Fig.\,\ref{fig:sigma_pcr_nondim} can be used to predict the critical
permeation pressure and breakup of emulsion droplets for specific
operating conditions and surface tension.

\subsection{The effect of contact angle on the droplet dynamics near the pore}
\label{subsec:contact_angle}

\begin{figure}[t]
\includegraphics[width=12.0cm,angle=0]{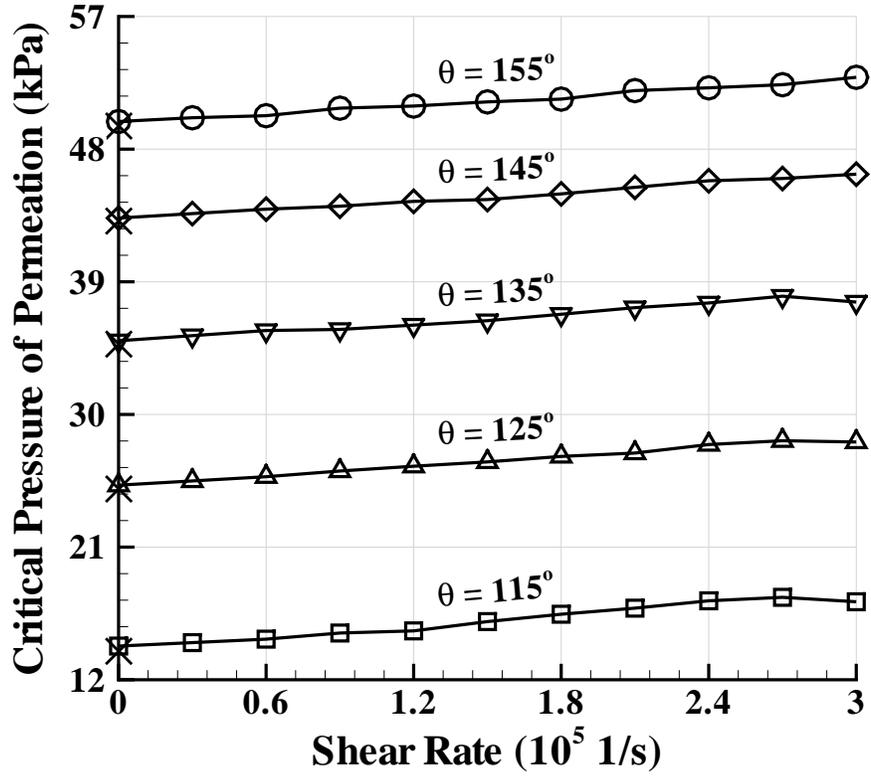}
\caption{The critical pressure of permeation as a function of the
capillary number for the indicated contact angles.   The critical
pressure at zero shear rate, given by Eq.\,(\ref{eq:pcritfull}), is
denoted by the symbols ($\times$).      The droplet radius, pore
radius, surface tension coefficient, and viscosity ratio are
$r_{d}=2\,{\mu}\text{m}$, $r_{p}=0.5\,{\mu}\text{m}$,
$\sigma=19.1\,\text{mN/m}$, and $\lambda=1$, respectively.  }
\label{fig:contact_angle_pcr}
\end{figure}

\begin{figure}[t]
\includegraphics[width=12.0cm,angle=0]{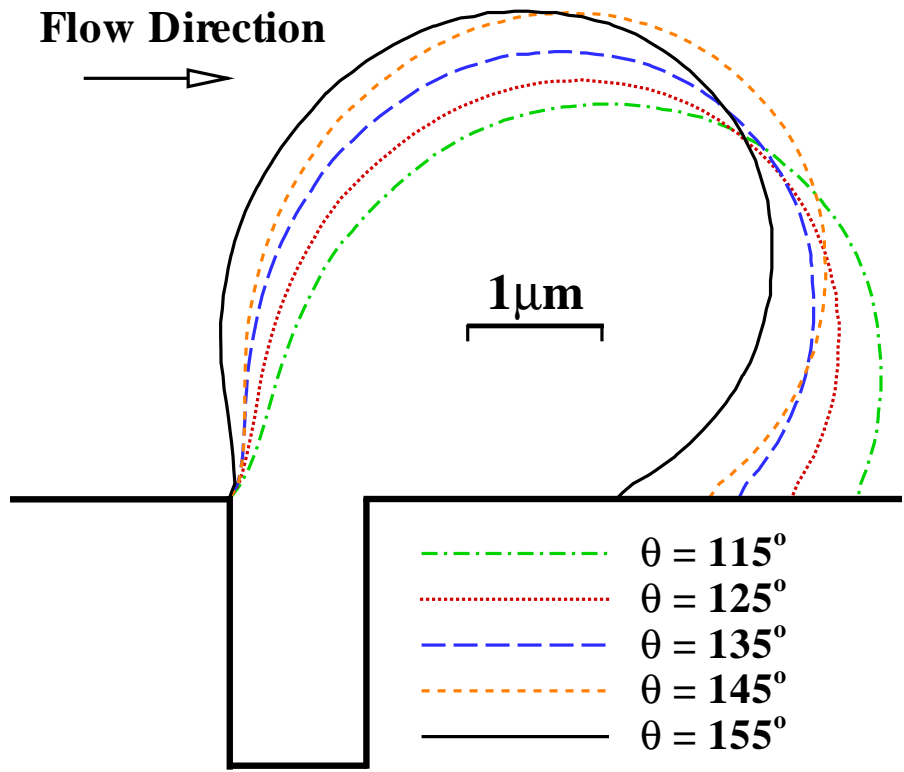}
\caption{The cross-sectional profiles of the oil droplet above the
circular pore for the listed values of the contact angle when
$Ca=0.022$.     Other parameters are the same as in
Fig.\,\ref{fig:contact_angle_pcr}. }
\label{fig:contact_angle_assembled}
\end{figure}

Next, we focus on the effect of contact angle on the permeation
pressure, deformation and breakup of oil droplets on a membrane
pore. The variation of the critical permeation pressure as a
function of the capillary number is presented in
Fig.\,\ref{fig:contact_angle_pcr} for nonwetting oil droplets with
contact angles $115^{\circ} \leqslant \theta \leqslant 155^{\circ}$.
The critical pressure at zero shear rate is higher for oil droplets
with larger contact angles, which is in agreement with the
analytical prediction of Eq.\,(\ref{eq:pcritfull}).    As expected,
with increasing shear rate, the critical pressure of permeation
increases for all values of $\theta$ studied.    We estimate the
maximum change in the critical pressure to be about 3\,kPa and
roughly independent of the contact angle.    This corresponds to a
relative increase of about $6\%$ for the contact angle
$\theta=155^{\circ}$ and $21\%$ for $\theta=115^{\circ}$.     These
results suggest that the relative efficiency of a microfiltration
system due to crossflow is higher for emulsion droplets with lower
contact angles.   Interestingly, we find that the critical capillary
number for breakup ($Ca_{cr}\approx0.032$) is nearly independent of
the contact angle.    This suggests that $Ca$ can be used as a
criterion for predicting breakup.   Finally, the examples of the
droplet cross-sectional profiles are shown in
Fig.\,\ref{fig:contact_angle_assembled}  for different contact
angles when $Ca=0.022$.   Notice that droplets with lower contact
angles wet larger solid area and are less tilted in the direction of
flow.

\subsection{The effect of droplet size on the critical pressure of permeation}
\label{subsec:droplet_size}

In the microfiltration process, the size of the membrane pore is one
of the crucial parameters that determine the permeate flux and
membrane selectivity.     Membranes with smaller pore sizes provide
higher rejections but require higher transmembrane pressures to
achieve the same permeate flux.   In this subsection, we examine the
influence of the drop-to-pore size ratio on the critical pressure of
permeation and the breakup dynamics of oil droplets in the presence
of crossflow above the membrane surface.

\begin{figure}[t]
\includegraphics[width=12.0cm,angle=0]{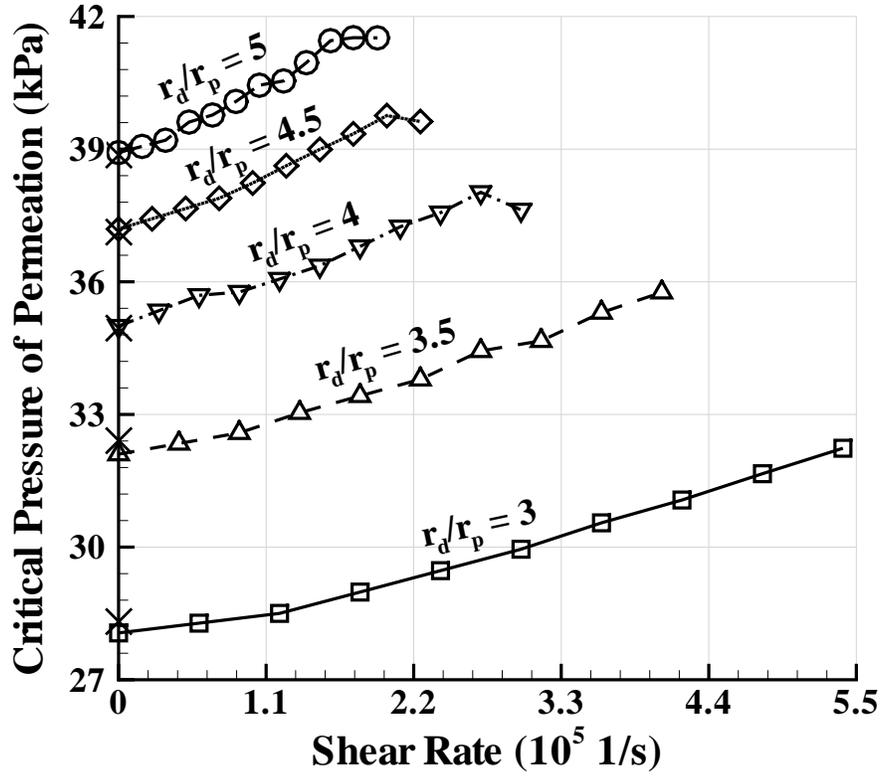}
\caption{The critical pressure of permeation as a function of shear
rate for the selected drop-to-pore size ratios.  The symbols
($\times$) indicate the critical pressure in the absence of flow
calculated from Eq.\,(\ref{eq:pcritfull}).  The pore radius, surface
tension coefficient, contact angle, and viscosity ratio are
$r_{p}=0.5\,{\mu}\text{m}$, $\sigma=19.1\,\text{mN/m}$,
$\theta=135^{\circ}$ and $\lambda=1$, respectively. }
\label{fig:rdrp_pcr}
\end{figure}

Figure\,\ref{fig:rdrp_pcr} reports the critical permeation pressure
as a function of shear rate for the droplet radii in the range from
$1.5\,{\mu}\text{m}$ to $2.5\,{\mu}\text{m}$, while the pore radius
is fixed at $r_{p}=0.5\,{\mu}\text{m}$.    In the absence of
crossflow, the critical pressure is higher for larger droplets
because they have lower curvature of the interface above the
membrane surface, which is in agreement with the analytical
prediction of Eq.\,(\ref{eq:pcritfull}).    With increasing shear
rate, the critical pressure increases for droplets of all sizes.
Note also that the slope of the curves in Fig.\,\ref{fig:rdrp_pcr}
is steeper for larger droplets because of the larger surface area
exposed to shear flow, resulting in a higher drag torque, and,
consecutively, a higher transmembrane pressure needed to balance the
torque.    Furthermore, as shown in Fig.\,\ref{fig:rdrp_pcr},
smaller droplets break at higher shear rates, since higher shear
stress are required to produce sufficient deformation for the
breakup to occur.   The maximum relative critical pressure is about
$14\%$ for $r_{d}/r_{p}=3$ and $6\%$ for $r_{d}/r_{p}=5$.

\begin{figure}[t]
\includegraphics[width=12.0cm,angle=0]{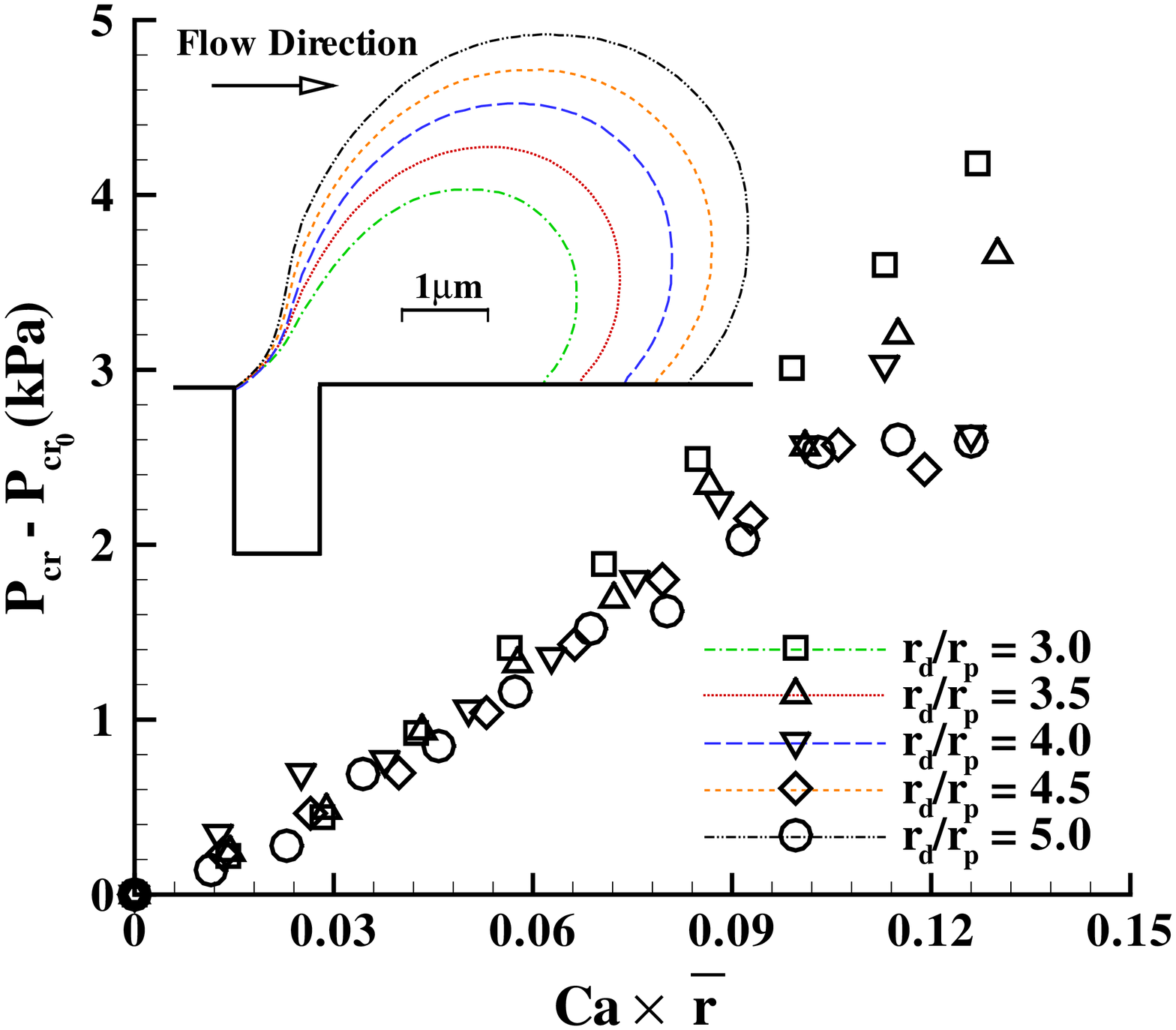}
\caption{The difference in the critical pressure,
$P_{cr}-P_{cr_{0}}$, versus the modified capillary number for five
drop-to-pore size ratios $\overline{r}=r_d/r_p$.     Other
parameters are the same as in Fig.\,\ref{fig:rdrp_pcr}.    The
cross-sectional profiles of the droplet just before breakup are
shown in the inset for same $\overline{r}$.     }
\label{fig:rdrp_assembled}
\end{figure}

We next compute the difference in the critical permeation pressure
with respect to the critical pressure in the absence of flow,
$P_{cr}-P_{cr_{0}}$, and define $\bar{r}=r_{d}/r_{p}$.    According
to Eq.\,(\ref{eq:ca_cr}), the product $Ca_{cr}\times\bar{r}$ is
independent of the droplet radius.     At the same time,
Eq.\,(\ref{eq:pressure_cr}) suggests that the increase in critical
pressure depends on the droplet radius via the term
$Ca\times\bar{r}$. Figure\,\ref{fig:rdrp_assembled} shows the
critical pressure difference as a function of the modified capillary
number $Ca\times\bar{r}$ for different droplet radii.        It can
be observed in Fig.\,\ref{fig:rdrp_assembled} that all curves nearly
collapse on each other and the droplet breakup occurs at the same
value $Ca\times\bar{r}\approx 0.125$.  We also comment that one of
the assumptions in deriving Eq.\,(\ref{eq:pressure_cr}) is that the
distance between the center of the pore and the center of the
droplet on the membrane surface is approximately $r_{d}$.    This
approximation becomes more accurate for larger drop-to-pore size
ratios, and, thus, the critical pressure difference in
Fig.\,\ref{fig:rdrp_assembled} is nearly the same for larger
droplets even at high shear rates.

The inset of Fig.\,\ref{fig:rdrp_assembled} shows the
cross-sectional profiles of oil droplets just before breakup for
different drop-to-pore size ratios.      Note that all droplets are
pinned at the pore entrance and elongated in the direction of flow.
It is seen that when $\bar{r}$ is small, the droplet shape is
significantly deformed from its original spherical shape.   In
contrast, larger droplets remain nearly spherical and only deform
near the pore entrance.    In general, the droplet-to-pore size
ratio should be large enough to make $P_{cr}$ sufficiently high for
practicable separation.  At the same time, if the pore size is much
smaller than the droplet size, the water flux through the membrane
decreases and the probability of breakup increases, which could
result in lower rejection rates and internal fouling of the
membrane. Therefore, choosing a membrane with an appropriate pore
size could greatly increase the efficiency of the microfiltration
process.

\section{Conclusions}

In this paper, we performed numerical simulations to study the
effect of material properties on the deformation, breakup, and
critical pressure of permeation of oil droplets pinned at the
membrane pore of circular cross-section.    In our numerical setup,
the oil droplet was exposed to a linear shear flow induced by the
moving upper wall.   We used finite-volume numerical simulations
with the Volume of Fluids method to track the interface between
water and oil.    The critical pressure of permeation was computed
using a novel procedure in which the critical permeation pressure
was found by adding pressure jumps across oil-water interfaces of
the droplet inside the pore and above the membrane surface. First,
the pressure jump across the static interface inside the pore was
calculated using the Young-Laplace equation. Then, the pressure jump
across the dynamic interface above the membrane surface was computed
numerically and added to the pressure jump inside the pore.  This
method has proven to be accurate, robust, and computationally
efficient.   To determine the dimensions of the computational
domain, we also studied the effect of confinement on the droplet
deformation and breakup and concluded that in order to minimize
finite size effects and computational costs, the distance between
the membrane surface and the upper wall has to be at least twice the
droplet diameter.  In particular, it was observed that highly
confined droplets become significantly deformed in a shear flow and
break up more easily.

In the absence of crossflow, we found that the analytical prediction
for the critical permeation pressure derived by Nazzal and
Wiesner~\cite{Nazzal96} agrees well with the results of numerical
simulations for different oil-to-water viscosity ratios, surface
tension, contact angles, and droplet sizes.    In general, with
increasing crossflow shear rate, the critical permeation pressure
increases with respect to its zero-shear-rate value and the droplet
undergoes elongation in the flow direction followed by breakup into
two segments.    The results of numerical simulations indicate that
at a fixed shear rate, the critical permeation pressure increases as
a function of the viscosity ratio, which implies that more viscous
droplets penetrate into the pore at higher transmembrane pressures.
In agreement with a scaling relation for the critical capillary
number, we also found that droplets of higher viscosity tend to
break at lower shear rates. Furthermore, with increasing surface
tension coefficient, the maximum increase in the critical permeation
pressure due to crossflow becomes larger and the droplet breakup
occurs at higher shear rates.   Interestingly, the percent increase
in critical permeation pressure as a function of the capillary
number was found to be independent of the surface tension
coefficient. Next, we showed that the breakup capillary number and
the increase in critical pressure of permeation are nearly
independent of the contact angle.   Last, it was demonstrated that
smaller droplets penetrate into the pore at lower pressures and
break up at higher shear rates because larger shear stresses are
needed to deform the interface above the membrane surface.

While most microfiltration membranes used in medium- to large-scale
separation applications have pores of complex morphologies and a
distribution of nominal sizes, results obtained for the simple case
of a pore of circular cross-section can be useful for identifying
general trends. With the development of new methods of manufacturing
micro-engineered membranes~\cite{Rijn13} and the rapid growth in the
diversity and scale of applications of microfluidic devices,
conclusions obtained in this work can be of direct practical value
for guiding membrane design and optimizing process variables.

\section*{Acknowledgments}

Financial support from the Michigan State University Foundation
(Strategic Partnership Grant 71-1624) and the National Science
Foundation (Grant No. CBET-1033662) is gratefully acknowledged.
Computational work in support of this research was performed at
Michigan State University's High Performance Computing Facility.



\bibliographystyle{prsty}

\begin{thebibliography}{99}


\bibitem{Morgan70}        J.T. Morgan, D.T. Gordon,
                          J. Petrol. Technol. 22 (1970) 1199.

\bibitem{Erdemir05}       A. Erdemir,
                          Tribology Int. 38 (2005) 249.

\bibitem{Lee01}           C. Lee, J. Lee, J. Cheon, K. Lee,
                          J. Environ. Eng. 127 (2001) 639.

\bibitem{Kaiser07}        H. Kaiser, N. Legner,
                          Plant Physiology, 143 (2007) 1068.

\bibitem{Brans04}         G. Brans, C.G.P.H. Schroen, R.G.M. van der Sman, R.M. Boom,
                          J. Membr. Sci. 243 (2004) 263.

\bibitem{Veil11}          J.A. Veil, Produced Water Management Options and Technologies.
                          In: Produced Water, Springer, 2011, pp. 537-571.

\bibitem{Baker13}         R. Baker, Membrane Technology and Applications. John Wiley and Sons, 2013.


\bibitem{Nazzal96}        F.F. Nazzal, M.R. Wiesner, Water Environ. Res. 68 (1996) 1187.

\bibitem{Cumming00}       I.W. Cumming, R.G. Holdich, I.D. Smith, J. Membr. Sci. 169 (2000) 147.

\bibitem{Hermia82}        J. Hermia, Trans. Inst. Chem. Eng. 60 (1982) 183.

\bibitem{Koltuniewicz95}  A.B. Koltuniewicz, R.W. Field, T.C. Arnot, J. Membr. Sci. 102 (1995) 193.

\bibitem{Mueller97}       J. Mueller, Y. Cen, R.H. Davis, J. Membr. Sci. 129 (1997) 221.

\bibitem{Gijsbertsen04}   A.J. Gijsbertsen-Abrahamse, A. van der Padt, R.M. Boom, J. Membr. Sci. 230 (2004) 149.

\bibitem{Walstra93}       P. Walstra, Chem. Eng. Sci. 48 (1993) 333.

\bibitem{Christopher07}   C.F. Christopher, S.L. Anna, J. Phys. D: Appl. Phys. 40 (2007) R319.

\bibitem{Vladisavl02}     G.T. Vladisavljevi\`{c}, S. Tesch, H. Schubert, Chem. Eng. Process. 41 (2002) 231.

\bibitem{Abismaill99}     B. Abisma\"{\i}l, J.P. Canselier, A.M. Wilhelm, H. Delmas, C. Gourdon, Ultrason. Sonochem. 6 (1999) 75.

\bibitem{Karbstein95}     H. Karbstein, H. Schubert, Chem. Eng. Process. 34 (1995) 205.

\bibitem{Bremond12}       N. Bremond, J. Bibette, Soft Matter 8 (2012) 10549.

\bibitem{Schramm92}       L.L. Schramm, Adv. Chem. Series 231 (1992).

\bibitem{Joscelyne00}     S.M. Joscelyne, G. Tr\"{a}g{\aa}rdh, J. Membr. Sci. 169 (2000) 107.

\bibitem{Lee84}           S. Lee, Y. Aurelle, H. Roques, J. Membr. Sci. 19 (1984) 23.

\bibitem{Hong03}          A. Hong, A.G. Fane, R. Burford, J. Membr. Sci. 222 (2003) 19.

\bibitem{Atencia05}       J. Atencia, D.J. Beebe, Nature 437 (2005) 648.

\bibitem{Puyvelde08}      P.V. Puyvelde, A. Vananroye, R. Cardinaels, P. Moldenaers, Polymer 49 (2008) 5363.

\bibitem{Stone94}         H.A. Stone, Annu. Rev. Fluid Mech. 26 (1994) 65.

\bibitem{Taylor32}        G.I. Taylor, Proc. R. Soc. Lond. A 138 (1932) 41.

\bibitem{Taylor34}        G.I. Taylor, Proc. R. Soc. Lond. A 146 (1934) 501.

\bibitem{Cox69}           R.G. Cox, J. Fluid Mech. 37 (1969) 601.

\bibitem{Hinch80}         E.J. Hinch, A. Acrivos, J. Fluid Mech. 98 (1980) 305.

\bibitem{Rumscheidt61}    F.D. Rumscheidt, S.G. Mason, J. Coll. Sci. 16 (1961) 238.

\bibitem{Grace82}         H.P. Grace, Chem. Eng. Commun. 14 (1982) 225.

\bibitem{Bentley86}       B.J. Bentley, L.G. Leal, J. Fluid Mech. 167 (1986) 241.

\bibitem{Husny06}         J. Husny, J.J. Cooper-White, J. Non-Newton. Fluid Mech. 137 (2006) 121.

\bibitem{Xu05}            J.H. Xu, G.S. Luo, G.G. Chen, J.D. Wang, J. Membr. Sci. 266 (2005) 121.

\bibitem{Rallison81}      J.M. Rallison, J. Fluid Mech. 109 (1981) 465.

\bibitem{Inamuro04}       T. Inamuro, T. Ogata, S. Tajima, N. Konishi, J. Comp. Phys. 198 (2004) 628.

\bibitem{Tryggvason01}    G. Tryggvason, B. Bunner, A. Esmaeeli, D. Juric, N. Al-Rawahi, W. Tauber,
                          J. Han, S. Nas, Y.-J. Jan, J. Comp. Phys. 169 (2001) 708.

\bibitem{Brackbill92}     J.U. Brackbill, D.B. Kothe, C. Zemach, J. Comp. Phys. 100 (1992) 335.

\bibitem{Gueyffier99}     D. Gueyffier, J. Li, A. Nadim, R. Scardovelli, S. Zaleski, J. Comput. Phys. 152 (1999) 423.

\bibitem{Scardovelli99}   R. Scardovelli, S. Zaleski, Ann. Rev. Fluid Mech. 31 (1999) 567.

\bibitem{Oneill68}        M.E. O'Neill, Chem. Eng. Sci. 23 (1968) 1293.

\bibitem{Price85}         T.C. Price, Q. J. Mech. Appl. Math. 38 (1985) 93.

\bibitem{Pozrikidis97}    C. Pozrikidis, J. Eng. Math. 31 (1997) 29.

\bibitem{Sugiyama08}      K. Sugiyama, M. Sbragaglia, J. Eng. Math. 62 (2008) 35.

\bibitem{Dimitra07}       P. Dimitrakopoulos, J. Fluid Mech. 580 (2007) 451.

\bibitem{Darvishzadeh12}  T. Darvishzadeh, N.V. Priezjev, J. Membr. Sci. 423-424 (2012) 468.

\bibitem{fluent}          Fluent, Inc., 2003. FLUENT 6.1 User's Guide.

\bibitem{Hirt81}          C.W. Hirt, B.D. Nichols, J. Comput. Phys. 39 (1981) 201.

\bibitem{Gerlach06}       D. Gerlach, G. Tomar, G. Biswas, F. Durst, Int. J. Heat Mass Transfer 49 (2006) 740.

\bibitem{Rider98}         W.J. Rider, D.B. Kothe, J. Comp. Phys. 141 (1998) 112.

\bibitem{Concus90}        P. Concus, R. Finn, Microgravity Sci. Technol. 3 (1990) 87.

\bibitem{Zapryanov98}     Z. Zapryanov, S. Tabakova, Springer, Vol. 50 (1998).

\bibitem{Renardy07}       Y. Renardy, Rheol. Acta. 46 (2007) 521.

\bibitem{Rallison84}      J.M. Rallison, Annu. Rev. Fluid Mech. 16 (1984) 45.

\bibitem{Stone88}         H.A. Stone, PhD Thesis, Caltech (1988).

\bibitem{Rijn13}          C.J.M. van Rijn, Micro-Engineered Membranes.
                          In: Encyclopedia of Membrane Science and Technology,
                          Eds: E.M.V. Hoek, V.V. Tarabara, John Wiley and Sons, 2013.

\end{thebibliography}

\end{document}